# Coupling of a quantum memory and telecommunication wavelength photons for high-rate entanglement distribution in quantum repeaters

Kyoko Mannami,[1,*] Takeshi Kondo,[1] Tomoki Tsuno,[1] Takuto Miyashita,[1] Daisuke Yoshida,[1] Ko Ito,[1] Kazuya Niizeki,[1,4] Ippei Nakamura,[3] Feng-Lei Hong,[1] and Tomoyuki Horikiri,[1,2]

[1] Department of Physics, Graduate School of Engineering Science, Yokohama National University, Yokohama 240-8501, Japan.

[2] JST PRESTO,4-1-8 Honcho, Kawaguchi, Saitama 332-0012, Japan.

[3] Komaba Institute for Science (KIS), The University of Tokyo, Meguro-ku, Tokyo, 153-8902, Japan.

[4] LQUOM Inc., Yokohama 240-8501, Japan.

*mannami-kyoko-kx@ynu.jp

**Abstract:** Quantum repeaters are indispensable tools for long-distance quantum communication. However, frequency matching between entangled photon sources and remote quantum memories (QMs) is difficult, which is an obstacle to the implementation of quantum repeaters. In this paper, we demonstrate a method to achieve the coupling of a Pr:YSO as a fixed-time QM with a single telecommunication-wavelength photon through frequency stabilization using an optical frequency comb over all applied laser wavelengths. The demonstrated method can lead to the implementation of a quantum repeater scheme enabling an improvement of the entanglement generation rate, paving the way for long-distance quantum communication.

# 1.Introduction

Modern telecommunication commonly involves the use of the Rivest–Shamir–Adleman (RSA) cryptosystem to ensure computational security [1]. However, ongoing improvements in high-performance computation threaten to compromise conventional security systems, making it imperative to develop new secure communication approaches. Quantum communication can fulfill this role by providing information-theoretically secure communication through quantum key distribution [2], as well as quantum secure direct communication which can transmit the secret message without sharing a key first [3-6]. An additional promising application of quantum communication is the quantum internet, which would allow for the worldwide connection of quantum devices via quantum state transfer. An example of this is blind quantum computation [7], which requires that users be able to transmit quantum states to a quantum computer server. In this manner, the quantum internet would serve as a platform that enables the use of various quantum technologies without users having to own high-performance quantum devices.

Establishment of quantum internet will require the ability to connect distant nodes forming the network, including end users, via quantum communication. Similar to the current internet, the quantum internet would be based on optical fiber network, but fiber loss is a significant problem for long-distance quantum communication. Even at the wavelength with the lowest loss (~1.5 μm), the fiber loss is approximately 0.2 dB/km. In classical systems, long-distance communication is achieved by amplifying the light signal; in quantum communication, it is impossible to perform this procedure owing to the no-cloning theorem. To address this issue, the use of quantum repeaters [8-12] to enable long-distance quantum communication has been proposed. Quantum repeater operations are performed using the following procedure: First, two photons in an entangled state are sent to two adjacent nodes, allowing these two parties to share the entangled state. The same procedure is then performed between all adjacent nodes. Quantum states that arrive at each repeater node are injected into Bell state analyzers to finally enable sharing of the entangled state between the end nodes by using a so-called nested protocol [8], thereby enabling the exchange of quantum information. In addition, entanglement purification is required when there is a decrease in entanglement fidelity in the process of increasing distance [13-15].

Recently, impactful studies on multiplexed quantum repeaters have been reported [16,17]. One study [16] realized single excitation entanglement (|01>+|10>), while another study achieved [17] two excitation entanglement (|HH>+|VV>) between two remote quantum memories. Both these studies used rare-earth doped materials, which are promising in terms of time division multiplexing [18] and wavelength division multiplexing [19]. These studies have realized entanglement sharing between separate quantum memories. The next step is to develop a system that can increase the entanglement distribution rate, which is an advantage of multiplexed communication, and can realize quantum communication mediated by one or more repeater nodes. In this paper, we report on the realization of a quantum device-coupling system to realize a method that can improve the two excitation entanglement distribution rate [10,20].

The remainder of this paper is structured as follows. Section 2 describes our proposal system. Section 3 describes the details of the experimental setup. Section 4 describes the results of our experiment to produce quantum memory (QM) using $Pr^{3+}$ doped $Y_2SiO_5$ (Pr:YSO) QM, which is coupled to telecommunication wavelength light. Finally, in Section 5, we discuss the possibility of applying the developed system to remote quantum device couplings and present the conclusions.

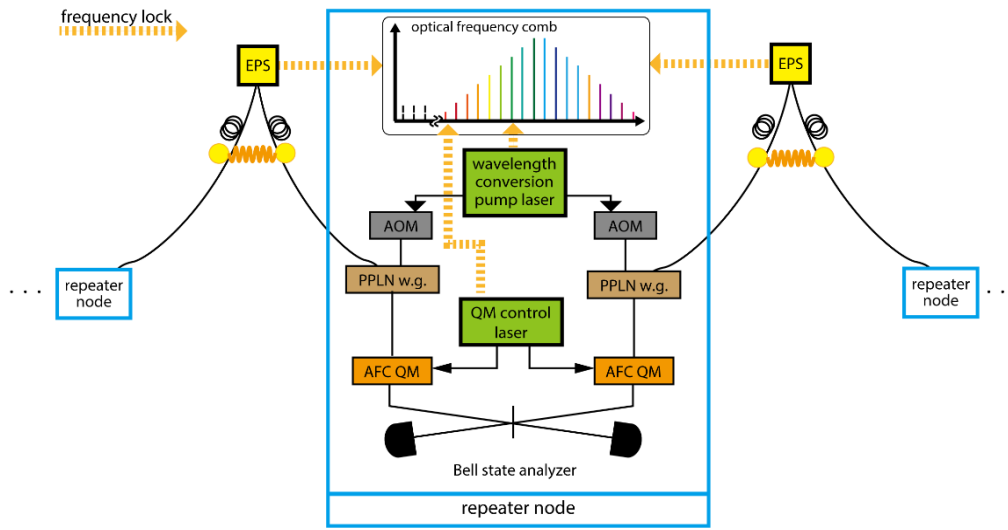

Fig. 1 Proposed quantum repeater protocol. Quantum entangled photon source (EPS) sends telecommunication wavelength entangled photons (Bell state which can be encoded in, for example, time-bin) through optical fibers to repeater nodes located on both sides. the telecommunication wavelength pump laser light is partially split from EPS and sent to an optical frequency comb installed in each repeater node, to stabilize its frequency. Upon arriving at each repeater node, the telecommunication wavelength photons from EPSs undergo wavelength conversion in periodically poled lithium niobate waveguide (PPLN.w.g.) and are absorbed and stored in the atomic frequency comb quantum memory (AFC QM). Wavelength-conversion pump laser and quantum memory control laser (QM control laser) light are also frequency-stabilized by the optical frequency comb in the repeater nodes. In the QMs located in each repeater node, the detunings of the wavelength-converted photons and QM control lasers are finely compensated using acousto-optic modulators (AOMs), respectively. Regenerated photons from the two QMs acquire the indistinguishability, making it possible to carry out a quantum entanglement swapping via a Bell state analyzer

## 2. System description

The quantum repeater scheme proposed in this paper is shown in Fig. 1. As the goal of quantum repeaters is to share quantum entanglement between two arbitrary points, the role of the proposed quantum repeater is to share this entanglement between adjacent repeater nodes, extend the distance via entanglement swapping, and increase entanglement fidelity through entanglement purification [8]. The system shown in Fig. 1 involves the entanglement of two photons emitted from entangled photon sources (EPSs) distributed between adjacent links. They are sent to the two sides of a communication channel and stored in the QMs of two quantum repeater nodes. Once the quantum states have been preserved by the QMs in the node, entanglement swapping is performed. The EPSs can be located either right next to the QM in the repeater node or between repeater nodes [9,10,20]. In this study, we used the latter configuration; however, the same method can be applied to the former.

Each EPS generates two photons at the telecommunication wavelength, with the two-photon generation performed via spontaneous parametric down-conversion (SPDC) in a nonlinear medium [21]. At this time, a narrow-linewidth pumping laser with a telecommunication wavelength (at approximately 1.5 μm) performs second-harmonic generation (at approximately 750 nm), which is used to pump the SPDC nonlinear medium, producing two photons with the same wavelength as the pumping laser. Each EPS can be a cavity-enhanced narrow linewidth two-photon source that increases coupling efficiency with the QM [22,23].

The telecommunication-wavelength entangled photons are sent to the quantum repeater nodes through optical fibers. Upon arriving at a repeater node, the quantum state of photons is transferred to the node's QM. Although the entangled state can have various degrees of freedom, time-bin entanglement is promising owing to its robustness to long-distance fiber transmission. An atomic frequency comb (AFC) [18] method using rare earth ion-doped crystals is suitable for the QM, which accepts time-bin signal. Previous studies on QMs used various rare earth ion-doped crystals [16,17,24-27]. Pr:YSO has relatively long coherence times, high multiplicities, and efficiencies.[28]. Because the transition wavelength for Pr:YSO is approximately 606 nm (Fig. 2 (a), [29]), the telecommunication wavelength photon must experience frequency conversion before entering the memory. A nonlinear optical process can be applied to achieve wavelength conversion with an extremely high conversion efficiency (in this case, a sum frequency generation (SFG) process of a telecommunication photon and a wavelength conversion pump laser (=1.0 μm) is applied to obtain a photon at 606 nm).

As shown in Fig.2(a) and 2(b), the AFC region of Pr:YSO is limited by the separation between the excited state hyperfine levels (~ 5 MHz). The frequencies of the telecommunication wavelength photon and the wavelength conversion pump laser must be stabilized within this region. Naturally, AFC generation (QM control) laser, which sets the working wavelength for the SFG light, must also be stabilized. In the proposed scheme, the frequency of SFG light is slightly different from that of the QM control laser. An acousto-optic modulator (AOM) is used to precisely match the frequencies. The difference between the frequencies can be set to zero via frequency modulation of the wavelength conversion pump laser by the AOMs shown in Fig. 1.

As shown in Fig. 1, each quantum repeater node contains two QM. After two photons from different EPSs have been loaded into the respective memories, they are reproduced as photons and used to perform entanglement swapping [30]. To obtain successful entanglement swapping via quantum interference in this process, the photons regenerated from the two QMs must be indistinguishable with both their timings and spectral (center frequency and line width) [31] matches when they are incident on the Bell state analyzer. In our scheme, each photon arrives at a repeater node probabilistically, therefore, timing information of photon arrival is necessary in the actual repeater implementation. AFC memory can have an adjustable storage time, however, heralding of photon arrival is necessary for two photons to enter the Bell state analyzer synchronously. In the full implementation of a quantum repeater, heralding which can be achieved with, for example, nondestructive photon arrival detection [32] is further needed.

As an optical frequency comb is placed in the repeater node, it is possible to make the frequencies of the telecommunication wavelength photons on both sides equal. Because a small amount of the telecommunication wavelength laser light can be transmitted via an optical fiber, this equalization can be performed with low amounts of loss. Because one wavelength conversion pump laser and one QM control laser suffice for two QMs, the problem of spectral matching between the regenerated photons produced by the two QMs can be avoided. Under this scheme, if frequency matching among the telecommunication wavelength laser, wavelength conversion pump laser, and QM control laser is stable for a sufficiently long period of time, coupling between EPSs and QMs and entanglement swapping are possible. In the following, we describe QM coupling with telecommunication wavelength photons via frequency matching and long-term frequency stabilization of all the involved wavelengths.

In this study, we used a telecommunication wavelength laser, whose intensity was attenuated to the single-photon level, as an alternative to an EPS. The feeble input stored by a QM was reproduced as a photon-echo and detected by a single-photon counting module. In previous QM coupling experiments involving wavelength conversion of single telecommunication wavelength photons [33,34], the QM control laser was generated via the SFG process of the telecommunication wavelength laser and a 1.0 μm conversion laser. Frequency matching was achieved between the photons and QM control laser because they were generated from the same light sources. In contrast, our approach adopts a more realistic system of independent lasers within the quantum repeater nodes, in which frequency adjustment between the lasers is important. To achieve this, we developed the necessary technology for matching the wavelengths of the wavelength-converted photon and QM. Thus, our research demonstrates a system that stably combines remote quantum systems (EPSs and QMs).

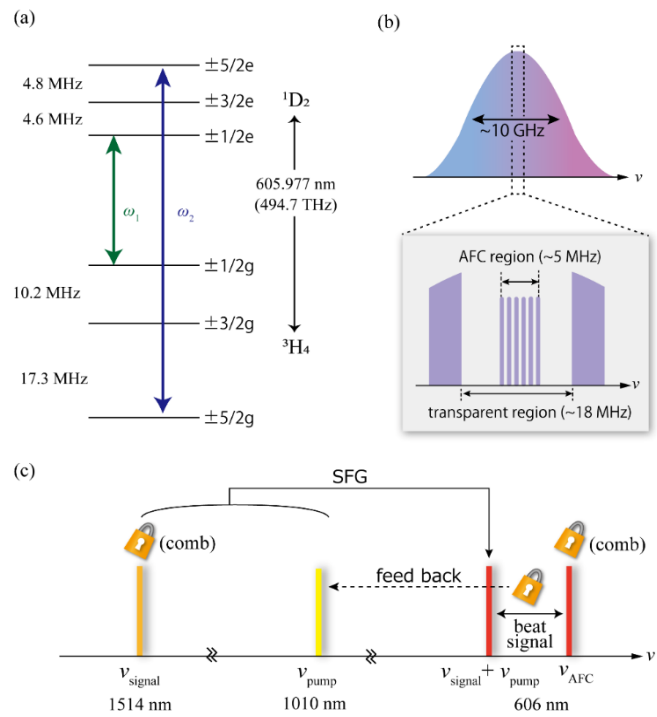

Fig. 2(a): Energy diagram of Pr:YSO . (b): Inhomogeneous broadening of Pr:YSO and atomic frequency comb (AFC) structure. (c) Wavelength conversion of telecommunication wavelength to QM wavelength using sum frequency generation (SFG). The lock marks correspond to frequency stabilizations. Details are discussed in the main text.

## 3. Experiment

The overall experimental system is shown in Fig. 3; each part is described below.

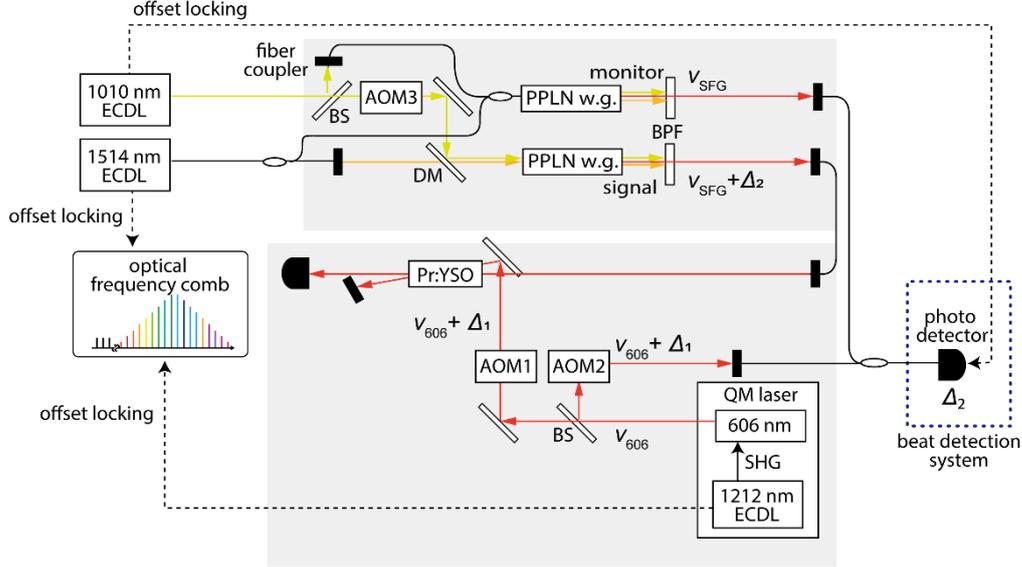

Fig.3 Experimental system. The solid black lines represent optical fibers. Sum frequency generation (SFG) of the telecommunication wavelength laser (1514 nm external cavity diode laser (ECDL)) and wavelength conversion pump laser (1010 nm ECDL) light occurs in the PPLN waveguides (PPLN w.g.s). In the upper PPLN w.g. path, monitoring is carried out to obtain a beat frequency with the QM control laser, with the result used to adjust the frequency via the AOM3. The signal from the lower PPLN w.g. is sent to the QM via an optical fiber. The signal is incident on the Pr:YSO crystal, in which an AFC structure has been created using a QM control laser, and photon echo signals are detected by the detector located behind the crystal. Several frequency stabilization processes are applied: the 1212 nm ECDL is a master laser of the QM control laser and offset-locked to the optical frequency comb; the 1514 nm ECDL is stabilized by offset-locking to the optical frequency comb, and the optical frequency comb synchronizes with GPS signals, which is the reference frequency in this study; the 1010 nm ECDL is offset-locked using the beat signal between the SFG light generated at the monitoring PPLN w.g. and the QM control laser modulated by the AOM2.

### 3.1 Fixed-time Quantum Memory

In the QM system, the AFC is created by causing optical excitations using a 606 nm laser (QM control laser). The laser power used to prepare the AFC is 1.5 mW. The excitation frequency and time are adjusted by applying a radio frequency signal to the AOM1 (double path configuration). To create the AFC, a transparent region is formed by sweeping the AOM1 frequency over an 18 MHz range centered at a frequency corresponding to 1/2g-1/2e and ω1 in Fig. 2 (a). Next, a single-frequency laser with a frequency corresponding to 5/2g-5/2e (ω2 in Fig. 2 (a)) is incident on the crystal to produce a sharp peak in the transparent region. By repeating this process at every comb interval (2 MHz in this case), an AFC is created. Total preparation time of AFC is about 500 ms. The Pr:YSO crystal used in the experiments had a $Pr^{3+}$ doping rate of 0.05% and dimensions of 1 mm × 1 mm × 5 mm, with a light polarization along $D_2$ axis[33]. We use a closed-cycle cryostat (Montana instruments, cryostation) to cool the sample to <4 K.

A 1514 nm laser converted to a wavelength of 606 nm is then incident on the memory. Using the AOM located in the 1010 nm light path (AOM3), the light becomes a pulse having a time width of about 90 ns (FWHM) and linewidth of approximately 4.5 MHz; the photon echo was observed by irradiating the light to the Pr:YSO crystal. The waiting time between preparation and storage is 20 ms. In the present study, the AFC is adopted as a fixed-time quantum memory [16,17]. However, as already proposed and demonstrated in several studies [33,35,36], quantum memory with a flexible storage time by transferring excited population onto a storage state can be obtained by using the same material.

### 3.2 Wavelength Conversion

1514 nm (197.9 THz) and 1010 nm (296.8 THz) lasers are incident on the PPLN waveguides (PPLN w.g.s) to produce 606 nm (494.7 THz) light via SFG. One of the two wavelength conversion systems is used for generating light to be stored in the memory (signal); the other is used for observing the beat signal for frequency matching (monitor). Using a band pass filter, only 606 nm light is transmitted and is carried by the fiber to the AFC QM. The 1010 nm (44 mW) and 1514 nm (730 µW) light incident on the PPLN w.g. produce a 606 nm light (509 µW), corresponding to a wavelength conversion efficiency of 27.9%.

### 3.3 Frequency Matching

As mentioned above, the key technique applied in the proposed method involves matching the frequency (wavelength) of the SFG signal to that of the QM. As our laboratory wavemeter had a resolution of 10 MHz and the actual QM frequency bandwidth was approximately 5 MHz, it was necessary to adjust the frequency with a higher accuracy.

The AFC is generated by modulating the light of the QM control laser (frequency of $\nu_{606}$) using AOM1. The center frequency of the modulation by AOM1 is $\Delta_1$; thus, the center frequency of the AFC can be written as $\nu_{606} + \Delta_1$. A part of the QM control laser is picked up by a beam splitter and modulated by AOM2. The modulation frequency of AOM2 was always set to $\Delta_1$, and this light is sent to the beat signal detection system. The SFG light from the monitor PPLN w.g. path also enters the beat detection system. The frequency of this light ($\nu_{SFG}$) is the sum of the frequencies of the 1010 nm and 1514 nm lasers. The 1010 nm laser is locked so that the beat signal frequency is $\Delta_2$, and the frequency relationship of the lasers is expressed as

$$\nu_{606}+\Delta_1-\nu_{SFG}=\Delta_2. \quad (1)$$

Because the SFG signal entering the memory, that is, the light from the signal PPLN w.g., is modulated by AOM3 with $f_{AOM3}$, the frequency can be expressed as $\nu_{SFG} + f_{AOM3}$. Our goal is to match the center frequency of the AFC with the frequency of the SFG signal incident on the AFC, i.e., to satisfy the

following relation

$$\nu_{606}+\Delta_1=\nu_{SFG}+f_{AOM3}. \qquad (2)$$

Comparing equations (1) and (2), we obtain $f_{AOM3} = \Delta_2$. In this experiment, we use 160 MHz and 164 MHz as $\Delta_1$, and $\Delta_2$, respectively. The frequency matching of the SFG signal and the QM is realized using the above scheme.

**3.4 Frequency Stabilization**

All the lasers used in the experiment are stabilized by the optical frequency comb using delay line based offset locking [37,38]. The 1514 nm laser and the 1212 nm laser (606 nm master laser) are stabilized by taking beat signals with the optical frequency comb, while the 1010 nm laser uses the beat signal, as mentioned in Section 3.3 (also shown in Fig. 2(c)).

As an example, Fig. 4 shows the stability of the beat signal between the monitoring signal (frequency = $\nu_{SFG}$) and QM control laser modulated by AOM2 (frequency = $\nu_{606}+ \Delta_1$). The results show the fluctuation of the SFG signal with respect to the center frequency of the AFC; the overall drift is within 150 kHz, which is less than 1/10 of the 2 MHz interval of the AFC, indicating that the stability required for the experiment has been obtained.

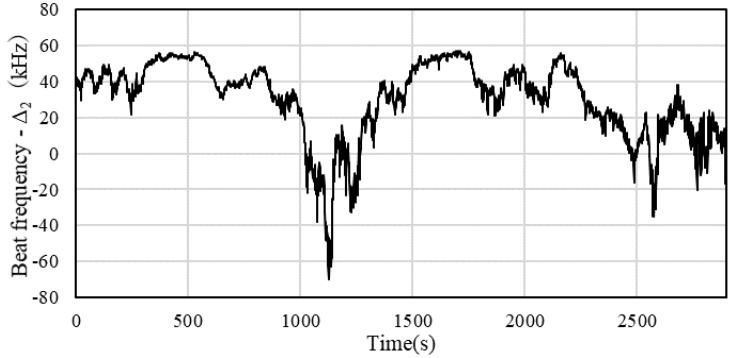

Fig. 4 Frequency stability of the beat signal between the monitor SFG and QM control laser. Black solid line shows the frequency variation of the beat signal (frequency = $\nu_{beat} - \Delta_2$). The fluctuation is less than 1/10 of the range determined by the AFC interval (dashed red lines).

## 4. Results

Figure 5 shows the generated AFC. The two peaks near 0 MHz are generated by the 1/2g-1/2e transitions, i.e., they indicate the AFC. The two peaks near 5 MHz correspond to the 1/2g-3/2e transition; these are not used in this study. In the photon echo experiment, a comb interval of approximately 2 MHz was used. From this, it would be expected that photon echo would have been observed at approximately 500 ns.

Figure 6(a) shows a photon echo obtained from a strong signal. The result in the figure reflects data obtained by eight repeated pulses (measurement time of less than 1 min) each time the AFC was generated and taking the average of the results observed by detecting the light transmitted through the AFC with a detector (Thorlabs, PDA10A2).

The black dotted line in Fig. 6 (a) shows the signal transmitted a transparent region (the AFC was not generated in this region); the red solid line shows the transmitted signal obtained when the AFC was generated in the region. In each case, the transmitted signal is observed near time = 0 ns. For the case in which the AFC was not formed, the signal can be considered to have been nearly transmitted. For the case in which the AFC was formed, some of the signal was absorbed and the signal energy at 0 ns was significantly reduced. An echo of this signal appears near 400 ns; this secondary signal does not occur in the signal produced when the AFC was not formed, thus confirming that echo represents temporal storage and reproduction of the incident signal. The reason why the echo emission time is earlier than the value expected from comb separation is that the number of AFC peaks is small and the fluctuation of the comb interval is large. In this case, the signal and echo energies are 3.92 and 0.31 pJ, respectively, corresponding to an echo efficiency of 7.9%. An additional signal appears at approximately 150 ns in the case in which the AFC formed. This signal is considered to be generated by the slow-light effect [39,40] induced by the gradient in the absorption peaks, i.e., the gradient of the refractive index in the crystal. Another possibility is caused by the overlap with the 1/2g-3/2e transitions since photon bandwidth determined by the input pulse is around 4.5MHz. In this case, the additional echo could be due to the beating of the outputs of the two combs (separated by ~5 MHz).

A similar experiment was then performed at the single photon level. The number of photons contained in the signal pulse was adjusted by installing a neutral density (ND) filter in the path of the 1514 nm laser. The averaged number of 1514 nm photons injected into the wavelength conversion crystal was 3.4 per 100 ns. After the light was pulsed by AOM3, signals containing an average of 0.59 photons per pulse were directed onto the AFC; this single photon echo experiment was performed 10,000 times over a measurement time of 15 min. The transmitted signals were coupled to a single-mode optical fiber, observed and detected using a single photon counting module (Perkin Elmer, SPCM-CD 3254), and then sent to a time interval analyzer. The results are shown in Fig. 6(b), which indicate that some part of the signal was absorbed by the AFC and then reproduced as an echo at approximately 400 ns. As in the case of strong light, the (unintentionally generated) slow light peak at 150 ns is also present.

The number of photons per echo was 0.040 per trial, corresponding to a photon echo efficiency of 6.8 %. The coincidence between the two cases in terms of the appearance and efficiency of echoes indicates that the proposed method ensures long-term stability of frequency tuning. When the power of the wavelength conversion laser is 40 mW, the noise generated from the wavelength conversion pump laser is $8\times10^2$ cps. This means that the noise per pulse is $8\times10^{-5}$, since the pulse width is about 100 ns. On the other hand, in the single-photon level experiment, the count of photons immediately after wavelength conversion is 0.96 per pulse (after including the coupling efficiency of the optical fiber after wavelength conversion and the loss before the quantum memory system, average photon number per pulse is reduced to 0.59). Therefore, the S/N is about $1\times10^4$, and the contribution of noise in this experiment is considered to be small.

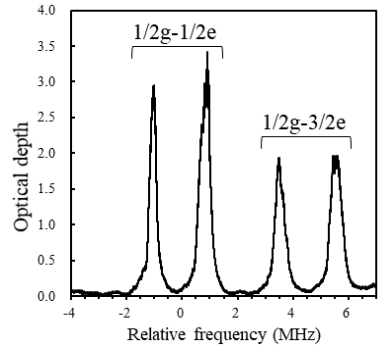

Fig. 5 Example of an AFC structure. Comb interval generated by the 1/2g-1/2e transition is 2 MHz. The pairs of peaks at approximately 5 MHz correspond to the 1/2g-3/2e transition to different excited states. We used only 1/2g-1/2e peaks as AFC.

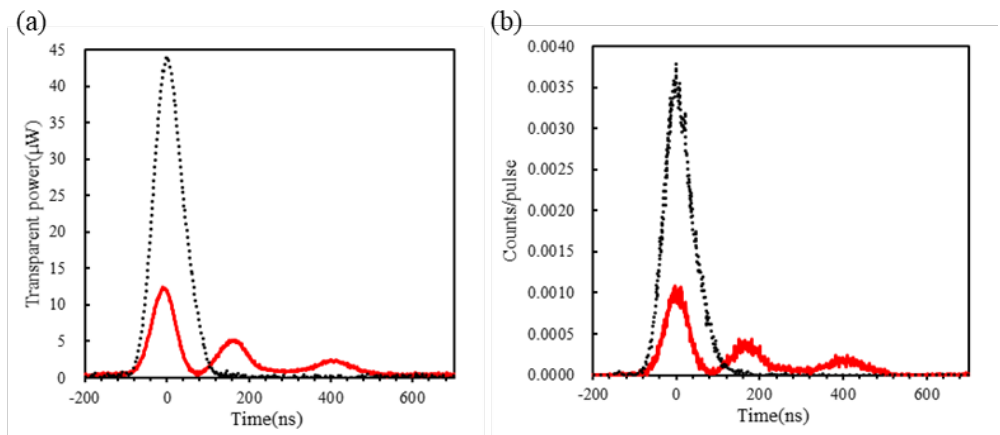

Fig. 6 Results of photon echo measurement using wavelength converted SFG signal. (a) strong signal and (b) single-photon level measurements.

## 5. Discussion and conclusion

In the proposed scheme, the wavelength of the telecommunication wavelength pump laser is the same as that of the entangled photon pair, enabling the development of a system of stably connected distant EPSs and QMs by using the following method. The proposed repeater generates a single AFC region in a Pr:YSO crystal to which single photons derived from a telecommunication wavelength laser are coupled through frequency matching. Locking the respective light sources with an optical frequency comb located in the repeater node is a direct approach, as it is sufficient to send a portion of a telecommunication wavelength pump laser through an optical fiber to provide effective locking with the optical frequency comb as shown in Fig.1. Another approach is that additional optical frequency combs locked to GPS signals are prepared close to EPSs. Although it is costly all the frequencies of the respective light sources can be well stabilized with the stability limited by the GPS combs.

Our goal is to couple the present system with QM having an arbitrary memory time. A highly stable narrow linewidth QM control laser is needed to obtain smaller comb separation, which enables longer AFC echo time and allows population transfer to the memory state for obtaining a longer storage time which will be bounded above by the coherence time of the nuclear spin states. In the present scheme, efficient population transfer is hampered by the poor spectral resolution of the presented AFC structure, which is limited by the linewidth of the QM control laser locked to the GPS comb. We need to prepare a frequency reference with a smaller frequency jitter than the GPS comb. To obtain narrower linewidths, we can explore the application of, for example, an optical frequency comb locking to a Nd:YAG laser with fast electronics [41]. The Nd:YAG laser can be locked by saturated absorption spectroscopy to the molecular iodine absorption line [42]. Using this method, it would be possible to obtain stabilities on the order of kHz or below through frequency stabilization with respect to the saturated absorption signal of the iodine line, resulting in a sufficiently long echo time.

The GHz bandwidth of the inhomogeneous absorption profile of Pr:YSO may be directly used to prepare a memory system [43]. However, it can be used for wavelength division multiplexing by forming a large number of MHz-order channels in Pr:YSO as shown in Ref [44]. The proposed method can also be used in its current form for frequency (wavelength) multiplexing communication. We are currently developing a two-photon telecommunication wavelength comb based on a cavity-enhanced entangled photon pair to take advantage of the numerous frequency modes with equidistant line widths of approximately 1 MHz (free spectral range (FSR) of approximately 100 MHz) [22]. As the frequency of the two-photon telecommunication wavelength comb is $v_{TPC} = v_{EPS} + mf_{FSR}$, where m is an integer and $v_{EPS}$ is the frequency of the telecommunication wavelength pump laser, large numbers of AFC structures can be generated at every FSR within the inhomogeneous broadening of Pr:YSO by performing wavelength division multiplexing [44]. In addition to wavelength division multiplexing, if we can make the AFC interval to be close to the homogeneous linewidth (~kHz), a large number of modes can be obtained for time division multiplexing. Specifically, the method described in the previous paragraph can be used. To generate AFC, it is necessary to irradiate a laser for a certain period of time, and the laser linewidth must be sufficiently narrow in order to obtain narrow AFC teeth. However, the linewidth is currently broader than 100 kHz due to the limitation of the GPS comb. To improve this, for example, a narrow-linewidth optical frequency comb stabilized to a Nd: YAG laser which is locked with a saturated absorption of molecular iodine line (to < kHz) can be achieved. If necessary, further reduction of linewidth can be achieved with locking the laser to a high finesse ultra-low expansion (ULE) cavity. Limitation of the AFC structure comes from excitation induced spectral diffusion [45] and/or inhomogeneity of the ground level spin-state transitions [46]. The limitation of tens of kilohertz in Pr:YSO set the upper bound to the multiplexing. For example, if a single photon from an EPS with 100 ns correlation time are injected into an AFC with 50 kHz interval, we can have $2 \times 10^2$ modes, since the echo time is 20 microseconds. In other words, the entanglement generation rate can be further improved by this amount. It is possible to increase the entanglement distribution rate by more than three orders of magnitude by using these multiplexing techniques.

To successfully implement a quantum repeater system, additional quantum entanglement purification and heralding functions must be introduced. However, the need for entanglement purification can be eliminated when the number of nodes is relatively small [10]. Additionally, heralding can be addressed through the use of a scheme in which a single photon from an entangled photon pair is absorbed just after emission by the memory in the repeater node and the other is sent through an optical fiber and enters a central station where Bell state measurement is performed [9,16,17] (in this case, one of the entangled photon pair is generated at the QM wavelength). In such a system, however, the entanglement generation rate might be relatively low, and it might not be possible to increase the number of repeater nodes. In the present case, the problem can be eliminated. The placement of EPSs at intermediate points between nodes [10,20] and a photon arrival's detection system [32] in repeater nodes could increase the entanglement generation rate several times in typical distances (tens of kilometers of node separation), enabling the use of the proposed scheme in remote long-distance quantum device coupling with guaranteed long-term operational stability.

In conclusion, we succeeded in stabilizing a telecommunication wavelength pump laser in a quantum repeater to the QM frequency via wavelength conversion and obtained photon echoes of single-photon signals. As discussed, the present quantum device coupling system can be extended to a system including multiple repeater nodes.

**Funding.** SECOM foundation; JST PRESTO (JPMJPR1769), JST START (ST292008BN), Japan Society for the Promotion of Science (JSPS) KAKENHI (JP20H02652).

**Acknowledgment.** We thank Shuhei Tamura and Kazumichi Yoshii for their support in the experiment.

**Disclosure.** The authors declare no conflicts of interest.